\documentclass[aps,pra,reprint,superscriptaddress,longbibliography,floatfix]{revtex4-2}

\usepackage{amsmath,amssymb,bm}
\usepackage{graphicx}
\usepackage{booktabs}
\usepackage{array}
\usepackage{hyperref}
\hypersetup{hidelinks}

\newcommand{\Tr}{\operatorname{Tr}}
\newcommand{\Frob}{\mathrm{F}}
\newcommand{\calP}{\mathcal{P}}

\newcommand{\calT}{\mathcal{T}}
\newcommand{\calC}{\mathcal{C}}
\newcommand{\calM}{\mathcal{M}}

\begin{document}

\title{Pulse-Informed Physical Placement for Parallel Two-Qubit Gates in Qubit Arrays}

\author{Shahid Qamar} 
\email{Corresponding to: shahidq@mail.ustc.edu.cn}  
\affiliation{Department of Mechanical Engineering, Sichuan University, Chengdu, 610016, China}
\affiliation{Beijing Academy of Quantum Information Sciences, Beijing 100193, China}
\author{Wei-Feng Zhuang}
\affiliation{Beijing Academy of Quantum Information Sciences, Beijing 100193, China}
\author{Bilal Riaz}
\affiliation{Department of Industrial and System Engineering, University of North Carolina, Charlotte, NC 28223, USA}
\author{Peng Qian} 
\email{Corresponding to: qianpeng@baqis.ac.cn}
\affiliation{Beijing Academy of Quantum Information Sciences, Beijing 100193, China}

\begin{abstract}
A parallel two-qubit layer can often be realized with several nonoverlapping sets of physical couplers, but isolated calibrations do not reveal how each realization performs after optimizing its own simultaneous control pulse. We formulate physical placement and pulse synthesis as a coupled decision problem for parallel controlled-$Z$ gates in heterogeneous frequency tunable transmon arrays. The proposed processor-aware structural quantum multi-domain matrix (QMDM) links static processor, edge, incidence, and task information to a dynamic candidate response block. For each candidate placement $P$, QMDM instantiates the active Hamiltonian $H_P$; quantum optimal control (QOCT) generates a placement-specific pulse, direct coupled propagation returns the projected computational subspace overlap $F_{\rm proj}$ and computational survival; and these pulse-conditioned responses are fed back to QMDM to rank and progressively promote physical placements. This placement pulse response feedback defines pulse-informed physical placement. Here, to see the performance of the proposed method on 30 held-out synthetic $3\times3$ processors, the median final $F_{\rm proj}$ is $0.993826$ for pulse-informed placement and $0.984981$ for calibration-based placement. Also, when tested under different system conditions, the proposed pulse-informed approach uses $59.1\%$ fewer candidate QOCT optimizer iterations than the exhaustive finite-budget reference. The QMDM representation and outer search are formulated for arbitrary $L$ simultaneous two-qubit gates. Here, the present global candidate ranking validation is used for $L=2$.
\end{abstract}

\maketitle

\section{Introduction}

Parallel quantum gates are the key components for system characterization~\cite{Erhard2019,Rudinger2021} and quantum error correction~\cite{Aharonov1997,Moore2001}. Parallel execution of two-qubit gates is required to limit circuit depth and make effective use of finite coherence times, but the quality of a gate measured in isolation does not generally determine its quality when neighboring operations are driven at the same time~\cite{Madzik2025}. Simultaneous gate characterization on superconducting processors has revealed correlated errors and substantial changes in gate performance under concurrent execution \cite{Rudinger2021,Zhao2022,Ketterer2023}.  While high-fidelity entangling operations with isolated qubit pairs were achieved early on, large-scale parallel gates have only recently been demonstrated on different platforms such as neutral atom arrays~\cite{Levine2019,Bluvstein2022,Evered2023}, superconducting circuits~\cite{Song2017,Cao2023,Chen2025,Fan2025}, and trapped ions~\cite{Figgatt2019}, but typically exceeding the fault-tolerant threshold~\cite{Arute2019,Wu2021}. Thus, realizing high-quality parallel gates is increasingly vital for leveraging quantum advantages on near-term devices~\cite{Preskill2018}.

Recent measurements further show that residual couplings across superconducting qubit arrays can possess pronounced spatial and spectral structure rather than being captured by a single uniform interaction scale \cite{Alghadeer2026}. These observations motivate treating simultaneous performance as a property of a physical configuration rather than as a simple product of isolated gate qualities. The issue is particularly important for frequency tunable transmons. At low energy, a transmon is well described as a weakly anharmonic multilevel oscillator \cite{Koch2007,Krantz2019,Blais2021}, and controlled phase gates can be implemented by moving computational states close to avoided crossings involving noncomputational levels. Adiabatic, interference assisted, and diabatic implementations all exploit this multilevel structure while controlling leakage and the accumulated conditional phase \cite{Martinis2014,Rol2019,GarciaRipoll2020,Barends2019,Negirneac2021}. The resulting gate quality depends on the local spectrum, intended coupling, nearby residual interactions, and the frequency trajectory actually applied during the gate. Hence, the achievable response of a physical edge is not fully specified by an isolated scalar calibration value.

At the compilation level, hardware-aware methods already exploit nonuniformity across a processor. Noise adaptive mapping uses calibration data and device variability to choose more favorable physical resources \cite{Murali2019}. Post-routing layout selection can enumerate isomorphic physical realizations and rank them with calibration derived heuristic costs, recovering a substantial fraction of the fidelity otherwise lost to unfavorable qubit selection \cite{Nation2023}. Frequency aware compilation and crosstalk-aware mapping or scheduling explicitly account for spectral crowding and simultaneous operation constraints \cite{Ding2020,CAMEL2025}. These methods establish that the physical realization of a logical circuit is itself an optimization variable.

Calibration information is nevertheless an imperfect predictor of circuit-level performance. Processor parameters drift, so recent calibration snapshots can become stale on experimentally relevant time scales; systematic studies of error-aware compilation therefore find that the benefit of calibration data depends on how and when it is used \cite{Kurniawan2024}. Context-based ranking can also outperform simple proxy scores: Hartnett \textit{et al.} showed that logically equivalent layouts on real hardware can differ strongly in performance and that a learned ranking model reduces layout selection error relative to both random and heuristic calibration based selection \cite{Hartnett2024}. At a larger scale, parallel CZ benchmarking has demonstrated inter-gate correlations that are invisible to independent local gate metrics and has used a global parallel-gate score directly as an optimization signal \cite{Fan2025}.

For quantum gate optimization, QOCT provides a systematic framework for designing continuous controls from a dynamical model \cite{Khaneja2005,Machnes2018,KochReview2022}. Pulse-level compilation similarly seeks to replace or aggregate gate-level instructions with hardware-specific controls \cite{Shi2019,AccQOC2020}. Crosstalk aware pulse and scheduling co-optimization shows that pulse design and simultaneous execution cannot always be treated as independent layers \cite{Xie2022}, while explicit crosstalk modeling and optimal control can recover high-quality parallel operations in systems with substantial local and nonlocal interactions \cite{Winick2021}. Calibration frameworks likewise combine control optimization, system characterization, and model refinement \cite{Wittler2021},Qian2025, recent closed-loop CZ optimization has reached $99.9\%$ fidelity with a 64 ns, seven-parameter Fourier pulse \cite{Glaser2025}.

For larger systems, subsystem-based QOCT addresses the exponential growth of the inner control problem by replacing full system robustness calculations with control problems on smaller subsystems while requiring robustness to inter-subsystem couplings \cite{Li2019}. A recent extension applies this construction explicitly to parallel gates and evaluates pair local robustness terms on constant size subsystem pairs, including demonstrations on superconducting qubit arrays \cite{Yang2026}. Existing methods usually separate physical placement from pulse optimization. Calibration-based methods rank candidate placements using static device information, while QOCT optimizes the control pulse after a placement has been chosen. However, for a fixed layer of \(L\) simultaneous two-qubit gates, several admissible physical placements may exist, and each defines a different Hamiltonian and control problem. Therefore, the best placement may only become clear after its candidate-specific pulse has been optimized and the resulting simultaneous response has been evaluated. This motivates treating physical placement and pulse optimization as a coupled problem.

The central observation of this work is therefore that physical placement and pulse synthesis are coupled. Let $P$ denote an admissible physical placement. Selecting $P$ fixes the active transmons and couplers and thereby defines a candidate Hamiltonian $H_P(t)$. That Hamiltonian determines the control landscape and the optimized pulse $u_P^{\star}(t)$. The pulse, in turn, determines how the local avoided crossings, leakage channels, and residual interactions are sampled during the gate. The simultaneous response is therefore a property of the placement together with its optimized control trajectory,
\begin{equation}
P\;\longrightarrow\;H_P
\;\longrightarrow\;u_P^{\star}(t)
\;\longrightarrow\;\left[F_{\rm proj}(P),S(P)\right].
\label{eq:conceptloop}
\end{equation}
Consequently, a ranking inferred from isolated edge calibration need not be preserved after candidate-specific simultaneous optimization.

We introduce a processor-informed structural quantum multi-domain matrix (QMDM) to represent and update this placement control relationship. QMDM is not a new pulse optimizer. Rather, it is the structured state of the outer placement problem: static blocks encode processor, edge, residual coupling, incidence, and task information, while a dynamic candidate block records the pulse-conditioned response of each evaluated physical realization. For a candidate $P$, QMDM supplies the parameters required to instantiate $H_P$; the inner QOCT problem generates a placement-specific pulse; direct coupled propagation returns $F_{\rm proj}$ and $S$; and those quantities are written back to the candidate block and used to update the placement ranking. We refer to this feedback as pulse-informed physical placement, because
the physical realization is selected using the dynamical response of a
pulse optimized for that realization itself.

The method retains isolated calibration as a low cost physical prior. A calibration policy commits to a placement $P_{\rm cal}$ using individually optimized edge qualities before simultaneous pulse dynamics enter the placement decision. The pulse-informed policy uses the same calibration data to construct a promising candidate region but keeps the placement decision open. Candidates then receive progressively increasing QOCT budgets, and their post-QOCT simultaneous responses determine which placements survive. Progressively allocating larger optimization budgets to a shrinking candidate is given in \cite{Jamieson2016,Hyperband2018}. Here, the allocation is embedded in a physics-based placement control loop in which each candidate is ranked using the response produced by its own optimized simultaneous dynamics.

\begin{figure*}[t]
\centering
\includegraphics[width=\textwidth]{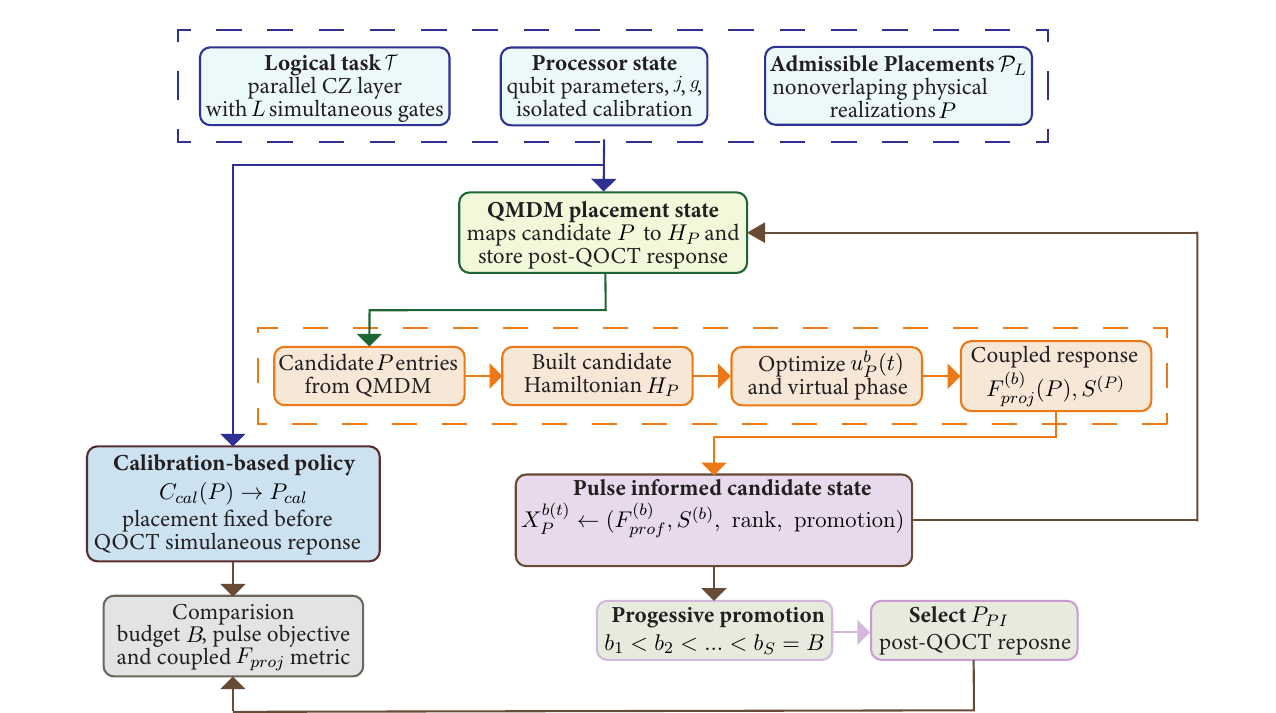}
\caption{\label{fig:architecture}
QMDM guided pulse-informed physical placement. The logical task, processor description, and admissible placement set define the static QMDM blocks. The calibration-based policy reads the isolated-edge information and fixes $P_{\rm cal}$ before simultaneous post-QOCT response is available. For the pulse-informed policy, a candidate $P$ is mapped through QMDM to its Hamiltonian $H_P$, optimized with QOCT at budget $b$, and evaluated by direct coupled propagation. The resulting $F_{\rm proj}^{(b)}(P)$ and $S^{(b)}(P)$ are written into the dynamic candidate response block, which is used to rank and promote placements. Repeating this feedback at increasing budgets produces $P_{\rm PI}$.}
\end{figure*}

We validate this framework on heterogeneous synthetic $3\times3$ frequency tunable transmon arrays for a fixed logical layer containing two simultaneous CZ gates. The study includes a 30-realization held-out ensemble, a separate processor on which all 44 admissible placements receive the common final QOCT budget, a stronger isolated calibration sensitivity test, numerical derivative and truncation checks, and an exploratory $L=3$ blockwise diagnostic. The placement representation and outer search are written for general $L$, whereas the exact global candidate ranking propagation used for the headline numerical results is demonstrated for $L=2$. Figure~\ref{fig:architecture} summarizes the information flow.

\section{Physical placement and gate model}

\subsection{Placement of a parallel two-qubit layer}

Let $G_H=(Q,E)$ denote the processor graph, where $Q$ is the set of physical transmons and $E$ is the set of native controlled-$Z$ capable edges. The logical task $\calT$ specifies $L$ two-qubit gates that must be executed in the same gate window. A physical placement is a set of $L$ nonoverlapping native edges,
\begin{equation}
P=\{e_1,\ldots,e_L\}\subset E,
\qquad e_k\cap e_j=\varnothing\quad(k\neq j),
\label{eq:placement}
\end{equation}
and the admissible placement set is
\begin{equation}
\begin{aligned}
\calP_L(\calT,G_H)=\{P\subseteq E:\;&|P|=L,\\
&e\cap e'=\varnothing\quad \forall e\neq e'\in P\}.
\end{aligned}
\label{eq:admissible}
\end{equation}
with any additional task-specific system constraints incorporated into $\calP_L$. In the benchmark, all requested operations are identical CZ gates, so an admissible realization can be represented by the unordered set of selected edges. If a logical layer contains labelled two-qubit operations with distinct mapping constraints, those labels and constraints belong to the task object $\calT$ and the admissible set is formed accordingly. The logical layer is fixed, but the elements of $\calP_L$ are alternative physical realizations. Only one placement is ultimately selected; the $L$ gates within that selected placement are executed simultaneously. Neither policy changes the logical gate sequence, serializes the layer, or reorders the requested operations in time. Throughout this paper, ranking or re-ranking refers only to sorting alternative physical placements according to a placement score; it does not refer to changing the logical order of the gates.

For a specified placement, the active Hamiltonian is written
\begin{equation}
H_P(t)=\sum_{k=1}^{L}\left[H_{S_k}+H_{C_k}(t)\right]
+\sum_{k<j}H_{S_kS_j},
\label{eq:Hparallel}
\end{equation}
where $S_k$ is the two-transmon subsystem associated with selected edge $e_k$, $H_{C_k}$ is its local control, and $H_{S_kS_j}$ contains residual interactions between distinct simultaneously active subsystems. Thus, changing $P$ changes the physical parameters entering $H_P$ and can change both the optimized pulse and the simultaneous gate response. This is also how spectral or frequency crowding enters the present model. Different placements select different values of $\omega_i$, $\alpha_i$, $J_e$, and cross-subsystem couplings $g_{ij}$; during a gate, the control excursion $u_k(t)$ moves the active transition frequencies and therefore changes the instantaneous detunings from the intended and unintended nearby avoided crossings. We do not introduce a separate static ``frequency-crowding score,'' independently retune the processor frequencies, or reorder the logical gate schedule. Instead, crowding among the modeled active transmons is represented dynamically in $H_P(t)$ and its effect is judged after candidate-specific pulse optimization through the directly propagated simultaneous response. This is consistent with the distinction between frequency-aware selection and pulse-level dynamical optimization in prior superconducting compilation and control studies \cite{Ding2020,CAMEL2025,Zhao2022}.

\begin{figure*}[t]
\centering
\includegraphics[width=0.96\textwidth]{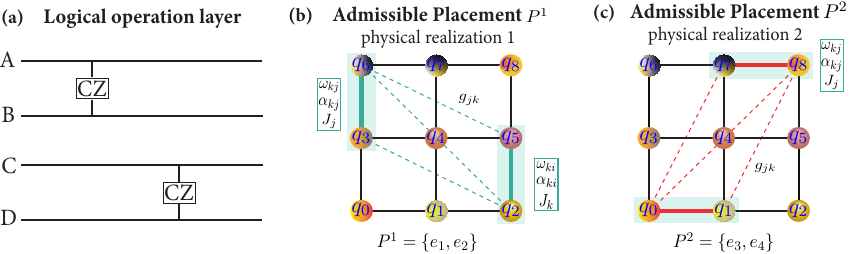}
\caption{\label{fig:placement}
Physical meaning of the placement variable. (a) The logical operation layer contains two simultaneous CZ gates and is fixed throughout the placement procedure. (b,c) Two admissible physical placements, \(P^{1}=\{e_1,e_2\}\) and \(P^{2}=\{e_3,e_4\}\), realize the same logical layer on different pairs of nonoverlapping native edges of the processor. Because the selected qubits and couplers differ, each placement is associated with different local frequencies \(\omega\), anharmonicities \(\alpha\), intended couplings \(J\), and residual inter-subsystem couplings \(g\). Consequently, the placements define different candidate Hamiltonians \(H_P\), generally lead to different optimized controls, and can produce different simultaneous gate responses.}
\end{figure*}

Figure~\ref{fig:placement} shows the representative placements used later in Sec.~V. The qubits of the $3\times3$ processor are numbered $q_0,\ldots,q_8$ in row major order. The calibration and pulse-informed choices share one native edge but use different partner edges for the second simultaneous controlled-$Z$ gate, making the physical meaning of the placement variable explicit.

\subsection{Three-level transmon model}

Each transmon is represented by the lowest three levels of the standard weakly anharmonic oscillator model \cite{Koch2007,Motzoi2009},
\begin{equation}
H_{\rm tr}=\omega a^\dagger a+\frac{\alpha}{2}a^\dagger a^\dagger aa,
\label{eq:transmon}
\end{equation}
where $\omega$ is the $0\leftrightarrow1$ angular frequency and $\alpha<0$ is the anharmonicity. The third level is required to represent the $|02\rangle$ and $|20\rangle$ channels relevant to a frequency-tuned controlled phase gate.

For selected subsystem $S_k$,
\begin{align}
H_{S_k}={}&\sum_{i=1}^{2}\left[\omega_{ki}a_{ki}^\dagger a_{ki}
+\frac{\alpha_{ki}}{2}a_{ki}^\dagger a_{ki}^\dagger a_{ki}a_{ki}\right]\nonumber\\
&-J_k(a_{k1}-a_{k1}^\dagger)(a_{k2}-a_{k2}^\dagger),
\label{eq:Hsub}\\
H_{C_k}(t)={}&u_k(t)a_{k1}^\dagger a_{k1},
\label{eq:Hcontrol}
\end{align}
and the residual interaction between two distinct active subsystems is
\begin{equation}
H_{S_kS_j}=-\sum_{i,l=1}^{2}g_{kj}^{il}(a_{ki}-a_{ki}^\dagger)(a_{jl}-a_{jl}^\dagger).
\label{eq:Hcross}
\end{equation}
The roles of $J$ and $g$ are distinct. $J_k$ is the intended coupling of the native edge selected as gate subsystem $S_k$, whereas $g_{kj}^{il}$ describes weaker interactions between transmons assigned to different simultaneously active subsystems. A selected native edge is therefore represented once, through its $J_k$ term, and cross-subsystem interactions enter through $g_{kj}^{il}$. Equations~\eqref{eq:Hsub}--\eqref{eq:Hcross} are an effective weakly anharmonic oscillator description of the active transmons, consistent with standard transmon control models \cite{Koch2007,GarciaRipoll2020}. Each native edge is oriented as ( high-band transmon, low-band transmon), and $u_k(t)\leq0$ denotes the instantaneous angular frequency shift applied to the high-band transmon. This effective frequency control model does not include the nonlinear flux-to-frequency transfer function or control line filtering and those hardware transfer effects are outside the present synthetic benchmark. Frequency-like parameters are reported divided by $2\pi$, while propagation uses angular frequency units.

For $L=2$, direct final verification propagates the four transmons belonging to the two active gate subsystems. The other processor qubits define the system graph and the admissible placement set but are not included as dynamical spectators. This modeling choice is revisited in Sec.~VII.

\subsection{Controlled-phase mechanism and performance metric}

\begin{figure*}[t]
\centering
\includegraphics[width=0.94\textwidth]{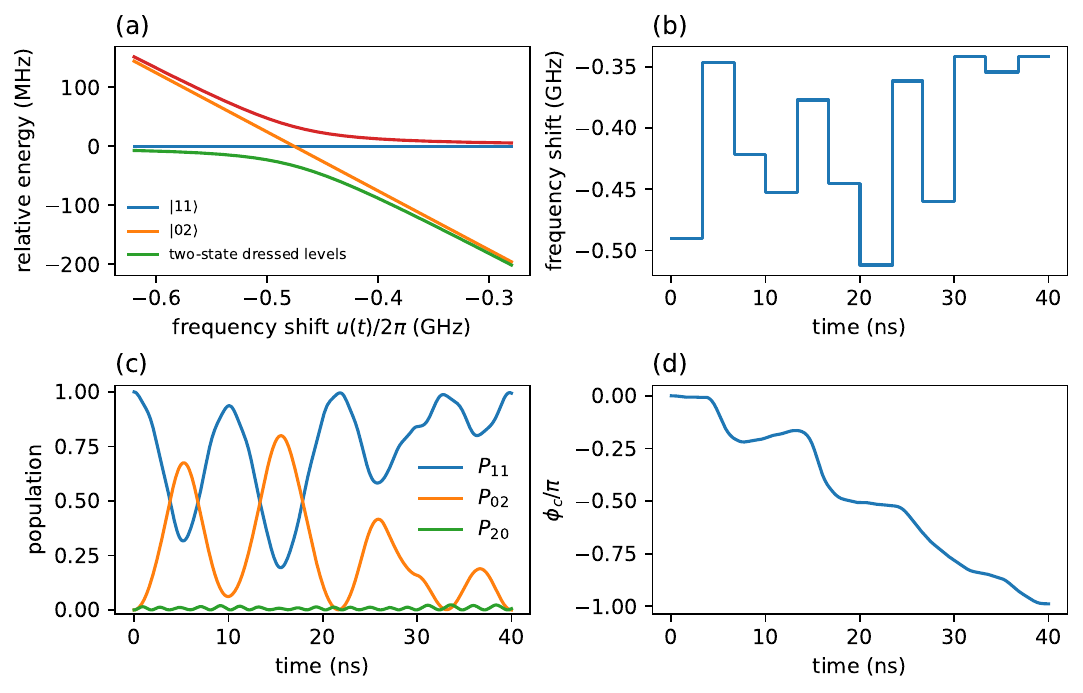}
\caption{\label{fig:cz}
Controlled-phase dynamics for one CZ subsystem of the representative pulse-informed placement. (a) Bare \(|11\rangle\) and \(|02\rangle\) energies as a function of the applied frequency shift \(u/2\pi\), together with the corresponding dressed eigenenergies, showing the avoided crossing used to generate the conditional phase. (b) Candidate-specific optimized frequency trajectory \(u(t)/2\pi\); the dashed line marks the bare \(|11\rangle\)-\(|02\rangle\) resonance. (c) Population evolution initialized in \(|11\rangle\), showing temporary transfer to the noncomputational \(|02\rangle\) state and subsequent return close to the computational subspace, while \(|20\rangle\) remains weakly populated. (d) Accumulated conditional phase \(\phi_c/\pi\), which approaches \(-1\), equivalent to the target CZ phase \(\pi\) modulo \(2\pi\). Together, the panels illustrate how the optimized frequency excursion exploits the avoided-crossing dynamics to accumulate the required conditional phase while recovering the computational population at the end of the gate.}
\end{figure*}

For one controlled-$Z$ subsystem, shifting the first transmon by $u(t)$ gives the bare two-excitation energies
\begin{align}
E_{11}(t)&=\omega_1+u(t)+\omega_2,\\
E_{02}&=2\omega_2+\alpha_2,\\
E_{20}(t)&=2[\omega_1+u(t)]+\alpha_1,
\end{align}
and hence the detunings
\begin{align}
\Delta_{02}(t)&=\omega_2+\alpha_2-[\omega_1+u(t)],\\
\Delta_{20}(t)&=\omega_1+u(t)+\alpha_1-\omega_2.
\end{align}
The coupling in Eq.~\eqref{eq:Hsub} connects $|11\rangle$ to $|02\rangle$ and $|20\rangle$ with matrix-element magnitude $\sqrt{2}|J|$. A controlled-$Z$ pulse approaches an accessible avoided crossing, allows a coherent excursion through the hybridized two excitation manifold, returns population to the computational subspace, and accumulates the conditional phase
\begin{equation}
\phi_c=\arg U_{11,11}-\arg U_{10,10}-\arg U_{01,01}+\arg U_{00,00},
\label{eq:condphase}
\end{equation}
with the target condition $\phi_c=\pi$ modulo $2\pi$. Figure~\ref{fig:cz} illustrates the mechanism for the representative pulse-informed solution.

The computational controlled-$Z$ target is $U_{\rm CZ}=\operatorname{diag}(1,1,1,-1)$. Local dynamical phases are corrected virtually using $Z(\phi)=\operatorname{diag}(1,e^{-i\phi})$, so the target for subsystem $S_k$ is
\begin{equation}
\bar U_{S_k}(\bm\phi_k)=
[Z(\phi_{k1})\otimes Z(\phi_{k2})]U_{\rm CZ}.
\end{equation}
If $P_{c,k}$ projects subsystem $S_k$ onto its computational basis and $U_{k,c}=P_{c,k}U_{S_k}(T)P_{c,k}^\dagger$, define
\begin{align}
f_{S_k}&=\frac{|\Tr(U_{k,c}\bar U_{S_k}^\dagger)|^2}{16},\\
s_{S_k}&=\frac{1}{4}\Tr(U_{k,c}^\dagger U_{k,c}).
\end{align}
For $L$ simultaneous gates,
\begin{equation}
P_C=\bigotimes_{k=1}^{L}P_{c,k},\qquad
\bar U(\bm\phi)=\bigotimes_{k=1}^{L}\bar U_{S_k}(\bm\phi_k),
\end{equation}
with $U_c=P_CU(T)P_C^\dagger$ and $d_c=4^L$. The final simultaneous metrics are
\begin{align}
F_{\rm proj}&=\frac{|\Tr(U_c\bar U^\dagger)|^2}{d_c^2},
\label{eq:Fproj}\\
S&=\frac{1}{d_c}\Tr(U_c^\dagger U_c).
\label{eq:survival}
\end{align}
Because leakage makes $U_c$ generally nonunitary, $F_{\rm proj}$ is a projected computational subspace overlap, not the standard average gate fidelity of a trace preserving quantum channel. Leakage and computational subspace survival are therefore tracked separately, consistent with the general distinction between in-subspace gate quality and leakage diagnostics \cite{WoodGambetta2018}. We report $S$ as the computational subspace survival probability averaged over computational basis states.

\section{QMDM guided pulse-informed placement}

\subsection{QMDM representation of processor and placement state}

The outer placement problem requires two kinds of information: quantities known before candidate pulse optimization and quantities learned only after a candidate has been driven by its own optimized pulse. We introduce QMDM as the structured state that carries these linked domains through the outer search. The term ``multi-domain matrix'' refers to the coupled matrix/table blocks and incidence maps; mathematically, $\calM$ is the following block collection rather than a single dense matrix:
\begin{equation}
\calM^{(b)}=\left\{X_Q,G_{QQ},X_E,B_{QE},C_{EP},X_P^{(b)},\calT\right\},
\label{eq:qmdm}
\end{equation}
where $b$ denotes the control budget used for the dynamic candidate response block. For $N_Q=|Q|$, $N_E=|E|$, and $N_P=|\calP_L|$, the processor side blocks are
\begin{align}
X_Q(i)&=(\omega_i,\alpha_i),\label{eq:XQ}\\
G_{QQ}(i,j)&=g_{ij},\label{eq:GQQ}\\
X_E(e)&=(J_e,F_e^{\rm cal},S_e^{\rm cal}),\label{eq:XE}\\
B_{QE}(i,e)&=\mathbf 1[i\in e],\label{eq:BQE}\\
C_{EP}(e,P)&=\mathbf 1[e\in P].\label{eq:CEP}
\end{align}
The qubit and edge identifiers, orientation, and high-/low-band labels are stored as metadata associated with these blocks. The task object $\calT$ specifies the requested gate type, the number $L$ of simultaneous gates, and any additional admissibility constraints.

For a proposed placement $P$, the incidence blocks identify its selected edges and qubits, $X_Q$ and $X_E$ provide the local transmon and intended coupling parameters, and $G_{QQ}$ provides the residual couplings between the simultaneously active subsystems. These entries instantiate the candidate Hamiltonian $H_P$ in Eq.~\eqref{eq:Hparallel}. In this sense QMDM is the state carried by the \emph{outer} placement problem: before pulse optimization it specifies what physical system corresponds to each candidate.

The block that makes the representation pulse-informed is $X_P^{(b)}$. After candidate $P$ has been optimized and propagated at budget $b$, we store
\begin{equation}
X_P^{(b)}(P)=
\left(F_{\rm proj}^{(b)},S^{(b)},
\operatorname{rank}^{(b)},\operatorname{promote}^{(b)}\right)_P.
\label{eq:XP}
\end{equation}
The first two entries are therefore not static system descriptors: they are pulse conditioned dynamical responses of the specific physical realization. This separation between the static processor blocks and the dynamic candidate response block is the central role of QMDM in the proposed method.

\subsection{Candidate pulse informed: QOCT and feedback}

This subsection describes the central operation of the method for one candidate placement. The outer search first proposes a physical placement $P$. QMDM then identifies the active qubits, native edges, frequencies, anharmonicities, intended couplings, and residual active couplings associated with that choice and uses them to instantiate $H_P$. At this point, the placement has not yet been judged by simultaneous pulse dynamics.

The inner QOCT gives the specified physical placement control pulse that best implements the requested simultaneous CZ layer. We parameterize the pulse by a finite set of control variables and virtual-$Z$ phases. For a piecewise constant segment $s$ of duration $\Delta t$, the candidate Hamiltonian is $H_{P,s}$ and the segment propagator is
\begin{equation}
U_{P,s}=\exp[-iH_{P,s}\Delta t],
\end{equation}
so that
\begin{equation}
U_P(T)=U_{P,N_s}\cdots U_{P,2}U_{P,1}.
\label{eq:segmentprop}
\end{equation}
Automatic differentiation is used to evaluate the gradient of the QOCT loss with respect to the pulse parameters and virtual phases, an approach that has been used directly for constrained optimal control problems in superconducting transmons \cite{Song2022}. Adam \cite{KingmaBa2015} performs the parameter updates. One numerical iteration therefore has the structure
\begin{equation}
\begin{aligned}
\text{pulse parameters}
&\rightarrow H_{P,s}\rightarrow U_P(T)\rightarrow \mathcal L_P,\\
\mathcal L_P
&\rightarrow \nabla\mathcal L_P\rightarrow \text{updated pulse parameters}.
\end{aligned}
\label{eq:qoctloop}
\end{equation}
The integer budget $b$ used below is simply the number of these optimizer updates. The specific pulse parameterization, segment count, initialization, learning rate, and benchmark budgets are given in Sec.~IV; they are numerical choices rather than part of the definition of pulse-informed placement. This is standard gradient-based numerical QOCT applied to a placement-dependent Hamiltonian \cite{Khaneja2005,KochReview2022,Machnes2018}.

After $b$ updates, QOCT returns the placement-specific pulse and virtual phases,
\begin{equation}
(u_P^{(b)},\bm\phi_P^{(b)})=\operatorname{QOCT}_{b}[H_P],
\label{eq:qoct}
\end{equation}
and direct propagation of the coupled active model gives
\begin{equation}
\mathcal R_b(P)=\left(F_{\rm proj}[P,u_P^{(b)}],\;S[P,u_P^{(b)}]\right).
\label{eq:response}
\end{equation}
These quantities are written back to the dynamic candidate block $X_P^{(b)}$. The placement has now acquired information that was unavailable from isolated calibration alone: its actual simultaneous response under a pulse optimized for that physical realization. The information flow is therefore
\begin{equation}
\begin{aligned}
(\calM,P)&\rightarrow H_P\rightarrow \operatorname{QOCT}_b\\
&\rightarrow (u_P^{(b)},\bm\phi_P^{(b)})
\rightarrow \mathcal R_b(P)\rightarrow X_P^{(b)}.
\end{aligned}
\label{eq:feedback}
\end{equation}
This feedback is the precise meaning of \emph{pulse-informed physical placement}. The logical gate layer has not been reordered; rather, alternative physical realizations of that same layer are re-ranked after each candidate has been allowed to reveal its own optimized simultaneous response.

To compare placement policies on the same final control problem, define a common final budget $B$ and
\begin{equation}
\mathcal F_B(P)=F_{\rm proj}[P,u_P^{(B)}].
\label{eq:commonfinal}
\end{equation}
The calibration-selected placement is optimized once at this final candidate QOCT budget after its location has been fixed. In the pulse-informed policy, the last promoted stage is defined with $b_S=B$; its final stage optimization both determines the winner among the finalists and supplies the reported $\mathcal F_B(P_{\rm PI})$. Hence the two reported final values use the same Hamiltonian model, pulse parameterization, objective, optimizer settings, and final budget. The difference is when pulse-conditioned simultaneous information is allowed to influence placement selection.

\subsection{Subsystem-regularized candidate QOCT}

The inner pulse optimization follows the subsystem-based robust control idea of Refs.~\cite{Li2019,Yang2026}. For every pair of simultaneously active gate subsystems $(k,j)$, define the crosstalk-free pair propagator
\begin{equation}
U_{0,kj}(t)=U_{S_k}(t)\otimes U_{S_j}(t),
\end{equation}
where $U_{S_k}$ is generated by $H_{S_k}+H_{C_k}(t)$. The first-order directional deviation associated with the residual coupling $H_{S_kS_j}$ is
\begin{equation}
D_{kj}=-iU_{0,kj}(T)
\int_0^T U_{0,kj}^\dagger(t)H_{S_kS_j}U_{0,kj}(t)\,dt.
\label{eq:Dkj}
\end{equation}
For the qutrit model, each pair calculation contains four transmons and has fixed dimension $d_{kj}=3^4=81$, independent of the total number $L$ of simultaneous gate subsystems. The benchmark optimizer uses a boundary-sampled approximation $\widetilde r_{kj}$ to the normalized squared directional sensitivity; its explicit form and numerical check are given in Appendix~\ref{app:validation}.

The pulse for candidate $P$ minimizes
\begin{equation}
\mathcal L_P=
1-\prod_{k=1}^{L} f_{S_k}
+\lambda_R\sum_{k<j}\widetilde r_{kj}
+\lambda_L\sum_{k=1}^{L}(1-s_{S_k})
+\lambda_S R_{\rm sm},
\label{eq:loss}
\end{equation}
with
\begin{equation}
R_{\rm sm}=\frac{1}{L(N_s-1)(2\pi)^2}
\sum_{k=1}^{L}\sum_{s=1}^{N_s-1}(u_{k,s+1}-u_{k,s})^2.
\end{equation}
The first term drives the intended CZ gates, the second reduces leading sensitivity to residual interactions between active subsystems, the third penalizes final leakage, and the fourth discourages abrupt segment to segment changes. Equation~\eqref{eq:loss} controls pulse synthesis only. It tells the inner optimizer how to improve the pulse for a fixed placement; it does not choose the placement itself. Candidate placement ranking is performed only after this inner optimization, using the directly propagated simultaneous $F_{\rm proj}$ of Eq.~\eqref{eq:Fproj}. Thus QOCT which pulse works for this $P$, whereas the outer QMDM search which $P$ should survive.

\subsection{Calibration baseline and progressive QMDM update}

Isolated-edge calibration solves a local CZ-control problem for each native edge without any other gate from the parallel layer. Let $F_e^{\rm cal}$ denote the resulting local quality. Using the static edge block $X_E$, define
\begin{equation}
C_{\rm cal}(P)=\prod_{e\in P}F_e^{\rm cal},
\label{eq:calscore}
\end{equation}
and
\begin{equation}
P_{\rm cal}=\arg\max_{P\in\calP_L}C_{\rm cal}(P).
\label{eq:pcal}
\end{equation}
This policy fixes the placement using only the pre-pulse information in the static QMDM blocks. It is intentionally an isolated calibration baseline: it asks which admissible placement contains the strongest individually optimized edges, before simultaneous post-QOCT response is known.

The pulse-informed policy starts from the same static information but keeps the placement decision open. Let
\begin{equation}
\mathcal A_K=\operatorname{Top}_K(\calP_L;C_{\rm cal})
\end{equation}
be $K$ calibration anchors. Using the edge placement incidence block $C_{EP}$, the first control evaluated candidate set is
\begin{equation}
\calC_1=\left\{P\in\calP_L:
\max_{A\in\mathcal A_K}|P\cap A|\ge \nu\right\}.
\label{eq:neighborhood}
\end{equation}
For the present $L=2$ layer, $\nu=1$ means that a candidate shares one selected native edge with at least one calibration anchor, so the search explores alternative partner edges around several strong isolated placements.

Let
\begin{equation}
b_1<b_2<\cdots<b_S=B
\end{equation}
be increasing candidate QOCT budgets and let $m_s$ denote the number of survivors after stage $s$. At stage $s$, every $P\in\calC_s$ is mapped through QMDM to $H_P$, optimized at budget $b_s$, directly propagated, and written back to $X_P^{(b_s)}$. The dynamic QMDM block is then used to retain
\begin{equation}
\calC_{s+1}=\operatorname{Top}_{m_s}
\left(\calC_s;F_{\rm proj}^{(b_s)}\right),
\qquad s=1,\ldots,S-1,
\label{eq:successive}
\end{equation}
and the final pulse-informed placement is
\begin{equation}
P_{\rm PI}=\arg\max_{P\in\calC_S}F_{\rm proj}^{(B)}(P).
\label{eq:select}
\end{equation}
Thus calibration proposes a physically promising region, but the placement is decided by pulse-conditioned simultaneous dynamics. In compact form,
\begin{equation}
\begin{aligned}
\text{placement}&\rightarrow\text{pulse}\rightarrow\text{coupled response}\\
&\rightarrow\text{placement ranking}.
\end{aligned}
\label{eq:pulseinformedbox}
\end{equation}
is the feedback loop that distinguishes $P_{\rm PI}$ from $P_{\rm cal}$.

The staged allocation avoids assigning the largest control budget to every admissible placement. For fresh optimization at each stage, the aggregate candidate QOCT iteration budget is
\begin{equation}
\mathcal B_{\rm PI}=\sum_{s=1}^{S}|\calC_s|b_s,
\label{eq:budgetPI}
\end{equation}
whereas evaluating every admissible placement at the final budget requires
\begin{equation}
\mathcal B_{\rm exh}=|\calP_L|B.
\label{eq:budgetexh}
\end{equation}
These are optimizer iteration counts used to compare search effort under the common $L=2$ candidate model; they are not wall-clock complexity laws. The benchmark values are specified in Sec.~IV.

\section{Numerical protocol}

\subsection{Synthetic processor ensemble}

We instantiate the method on synthetic $3\times3$ nearest-neighbor transmon arrays with heterogeneous frequencies, anharmonicities, native couplings, and residual interactions. The ranges below define a controlled numerical stress test rather than a fit to one particular fabricated processor; their scale is chosen to remain representative of weakly anharmonic transmon and residual coupling regimes discussed in the superconducting-control literature \cite{Koch2007,Rol2019,Alghadeer2026,Yang2026}. Qubits are numbered $q_0,\ldots,q_8$ row by row as in Fig.~\ref{fig:placement}. The processor and control parameter distributions are summarized in Table~\ref{tab:params}. The residual matrix is drawn once for each processor realization and then treated as the nominal model supplied to both placement policies. For this benchmark, the QMDM static blocks have dimensions $X_Q:(9,2)$, $G_{QQ}:(9,9)$, $X_E:(12,3)$, $B_{QE}:(9,12)$, and $C_{EP}:(12,44)$. The dynamic candidate response block has one row for each evaluated placement at a given budget. Thus the same QMDM construction used in Sec.~III is instantiated separately for every synthetic processor realization.

\begin{table}[t]
\caption{\label{tab:params}Synthetic processor and control parameters. Frequency-like quantities are reported divided by $2\pi$; Hamiltonian propagation uses angular frequency units.}
\begin{ruledtabular}
\begin{tabular}{lc}
Quantity & Value or distribution\\
\hline
High-band $\omega/2\pi$ & $\mathcal U(6.00,6.30)$ GHz\\
Low-band $\omega/2\pi$ & $\mathcal U(5.70,6.00)$ GHz\\
Anharmonicity $\alpha/2\pi$ & $\mathcal U(-0.275,-0.255)$ GHz\\
Native coupling $J/2\pi$ & $\mathcal U(0.022,0.026)$ GHz\\
Nominal residual scale $g_{\max}/2\pi$ & $1.5$ MHz\\
NN residual $g_{ij}/2\pi$ & uniform in $[-2g_{\max},2g_{\max}]$\\
Non-NN residual $g_{ij}/2\pi$ & uniform in $[-g_{\max},g_{\max}]$\\
Gate time $T$ & $40$ ns\\
Pulse segments $N_s$ & $12$\\
Allowed $u/2\pi$ & $[-0.8,0]$ GHz\\
\end{tabular}
\end{ruledtabular}
\end{table}

For the main study $L=2$, the $3\times3$ grid contains 12 native edges and 44 admissible nonoverlapping two-edge placements. The search settings are given in Table~\ref{tab:searchsettings}. These numbers define the benchmark; they are not part of the general method.

\begin{table}[b]
\caption{\label{tab:searchsettings}Numerical instantiation of the progressive placement search.}
\begin{ruledtabular}
\begin{tabular}{lc}
Quantity & Benchmark value\\
\hline
Processor topology & $3\times3$ nearest-neighbor grid\\
Parallel layer size & $L=2$ controlled-$Z$ gates\\
Native edges & $|E|=12$\\
Admissible placements & $|\calP_2|=44$\\
Calibration anchors & $K=3$\\
Structural overlap & $\nu=1$ shared edge\\
Progressive stages & $S=3$\\
Control budgets & $(b_1,b_2,B)=(40,100,160)$ iterations\\
Retained candidates & $(m_1,m_2)=(12,4)$\\
Held-out processors & 30 realizations\\
\end{tabular}
\end{ruledtabular}
\end{table}

\subsection{Control optimization and choice of search budgets}

Piecewise constant or low-dimensional parameterizations are standard in numerical quantum control \cite{Khaneja2005,KochReview2022}; here the bounded frequency controls are parameterized by unconstrained variables $v_{k,s}$ as
\begin{equation}
\frac{u_{k,s}}{2\pi}=-0.8\,\sigma(v_{k,s})\ \mathrm{GHz},
\qquad
\sigma(x)=\frac{1}{1+e^{-x}}.
\label{eq:controlparam}
\end{equation}
Candidate control uses 12 segments, Adam learning rate $0.025$, deterministic initialization $0.2\,\mathcal N(0,1)$ with numerical seed 123, and zero initial virtual-$Z$ phases. The loss weights in Eq.~\eqref{eq:loss} are $(\lambda_R,\lambda_L,\lambda_S)=(1,0.05,10^{-5})$.

Isolated-edge calibration uses eight pulse segments, 50 Adam iterations, learning rate $0.03$, and a fixed numerical initialization. Its local objective is
\begin{equation}
\mathcal L_e^{\rm cal}=1-F_e^{\rm cal}+0.05(1-S_e^{\rm cal})+10^{-5}R_{\rm sm}.
\end{equation}
No cross-subsystem regularizer appears in this calibration objective because only one gate subsystem is present; this is intentional and defines the calibration only baseline.

The progressive budgets and survivor counts in Table~\ref{tab:searchsettings} were fixed using development calculations before the 30 held-out processors were evaluated. The first budget was chosen to provide an inexpensive early control signal, but development cases showed that the short-budget ranking could be unreliable; 12 candidates were therefore retained rather than selecting the earliest winner. The intermediate budget refines this ordering, and the largest budget defines the final numerical test. The value $B=160$ is therefore a benchmark optimization budget, not a universal physical constant or a required iteration count of the method. On another optimizer or device model, the sequence $b_1<\cdots<B$ can be replaced by budgets chosen from convergence tests or a stopping criterion. In the present benchmark, candidates promoted to a larger budget are reoptimized from the same deterministic initialization rather than warm-started from the previous stage, so Eq.~\eqref{eq:budgetPI} counts the full budget at every stage.

All processor distribution parameters, control settings, search budgets, survivor counts, and statistical procedures were fixed before evaluating the 30 held-out realizations. Device-dependent quantities, including the calibration anchors, candidate-set size, optimized pulses, and selected placement, are recomputed independently for every processor.

\subsection{Numerical consistency checks}

The implementation was checked at the propagator, factorization, derivative, and model truncation levels. For a representative pulse segment, we evaluate the Frobenius unitarity residual $\|U_s^\dagger U_s-I\|_{\Frob}$, which is $5.1\times10^{-13}$. Setting all inter-subsystem couplings to zero gives a Frobenius difference of approximately $2.0\times10^{-12}$ between direct four-qutrit propagation and $U_{S_1}\otimes U_{S_2}$. The analytic piecewise-constant directional derivative agrees with a centered finite difference derivative of the complete propagator to relative error $1.7\times10^{-12}$. Additional checks of the boundary-sampled regularizer and the three-level truncation are reported in Appendix~\ref{app:validation}; these checks are used to validate the numerical implementation and do not enter the placement ranking.

\section{Results}

All placement scores in this section use the final quantity $\mathcal F_B$ of Eq.~\eqref{eq:commonfinal}. Isolated calibration enters the static QMDM edge layer and is used to select $P_{\rm cal}$ or initialize the pulse-informed candidate set; the dynamic QMDM candidate layer is populated only after candidate-specific QOCT and direct simultaneous propagation. The reported final comparison is therefore between two placement decisions evaluated with the same pulse model and metric. No logical gate sequence is changed anywhere in this section: the two CZ gates remain simultaneous. What changes is only the ranking of the admissible physical edge sets that could realize that fixed parallel layer.

\subsection{Representative example: pulse-induced re-ranking of physical placements}

The representative processor in Fig.~\ref{fig:placement} provides a concrete step by step example of the feedback loop. The fixed logical task is still the same pair of simultaneous CZ gates. What differs is which physical edges are used to realize that task. Isolated calibration selects
\begin{equation}
P_{\rm cal}=\{(q_2,q_1),(q_8,q_5)\},
\end{equation}
whereas the pulse-informed search eventually selects
\begin{equation}
P_{\rm PI}=\{(q_4,q_1),(q_8,q_5)\}.
\end{equation}
The two candidates share $(q_8,q_5)$ and differ only in the partner edge. Their isolated calibration products are $C_{\rm cal}(P_{\rm cal})=0.9437$ and $C_{\rm cal}(P_{\rm PI})=0.8973$, so the static calibration ranking places $P_{\rm cal}$ first and $P_{\rm PI}$ fifth among the 44 admissible physical placements. This is the ranking before simultaneous candidate-specific QOCT has been allowed to influence the decision.

\begin{figure*}[t]
\centering
\includegraphics[width=0.96\textwidth]{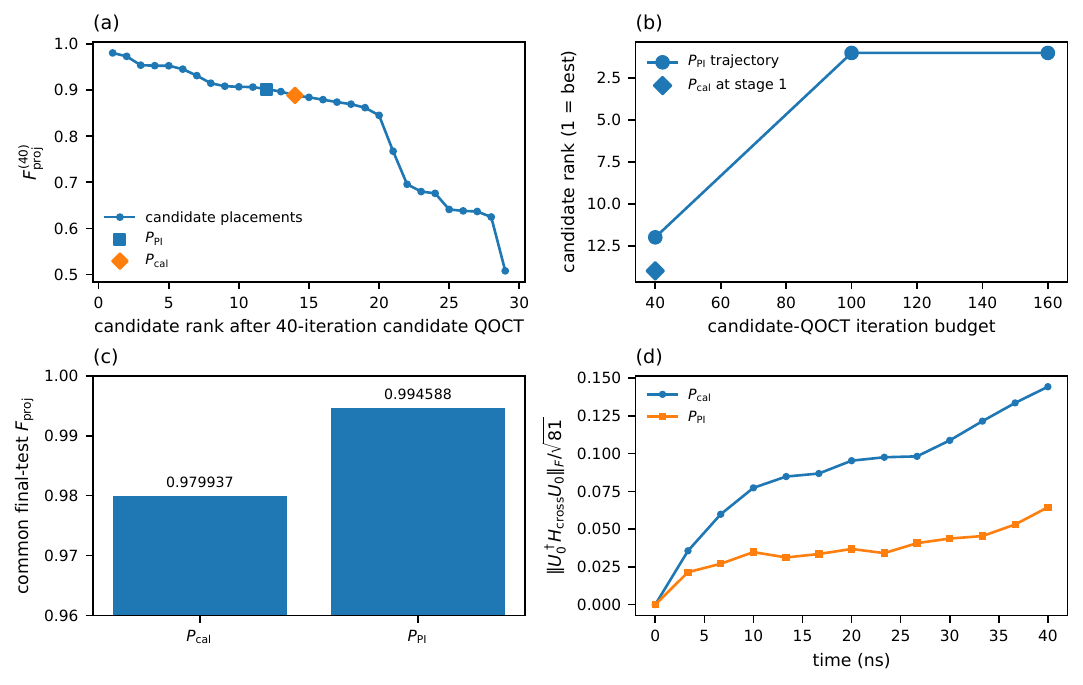}
\caption{\label{fig:representative}
Representative example of pulse-informed re-ranking of physical placements. The logical layer of two simultaneous CZ gates is fixed throughout; only the physical placement is changed. (a) Candidate ranking after the first finite QOCT budget \(b_1=40\). The calibration-selected placement \(P_{\mathrm{cal}}\) and the eventual pulse-informed placement \(P_{\mathrm{PI}}\) occupy different positions once candidate-specific simultaneous dynamics are included. (b) Evolution of the rank of \(P_{\mathrm{PI}}\) as the candidate-QOCT budget increases from 40 to 100 and 160 iterations. (c) Final comparison after \(P_{\mathrm{cal}}\) and \(P_{\mathrm{PI}}\) are evaluated using the same final QOCT budget and coupled system metric, although \(P_{\mathrm{cal}}\) is favored by isolated calibration, \(P_{\mathrm{PI}}\) gives the larger final projected overlap. (d) Interaction picture deviation
\(\delta(t)\)for the final optimized controls. The smaller deviation for \(P_{\mathrm{PI}}\) indicates weaker pulse-dressed residual interaction effects for this representative processor.}
\end{figure*}

The pulse-informed search then asks how these physical realizations behave after their own control problems are solved. At the first budget $b_1=40$, the eventual $P_{\rm PI}$ is rank 12 within the initial control-evaluated candidate set, so the short optimization is not yet trusted as a final decision. The 12 best candidates are retained. When those survivors are reoptimized from the common initialization at $b_2=100$, the eventual $P_{\rm PI}$ rises to rank 1. It remains rank 1 when the four finalists are evaluated at $B=160$. Hence Fig.~\ref{fig:representative}(a,b) does not show a change in gate order; it shows a change in the \emph{ordering of physical-placement candidates} as more reliable pulse-conditioned information becomes available. The candidate that was only fifth under isolated calibration becomes the best surviving physical realization after its own optimized simultaneous dynamics are taken into account.

At the common final budget,
\begin{align}
\mathcal F_B(P_{\rm cal})&=0.979937,& S(P_{\rm cal})&=0.993646,\\
\mathcal F_B(P_{\rm PI})&=0.994588,& S(P_{\rm PI})&=0.996641.
\end{align}
To examine the mechanism, define
\begin{equation}
\delta(t)=\frac{\|U_0^\dagger(t)U(t)-I\|_{\Frob}}{\sqrt{81}},
\label{eq:delta}
\end{equation}
where $U_0$ is the crosstalk-free product evolution and $U$ is the coupled four-qutrit evolution under the same controls. At $T$, $\delta\approx0.1443$ for $P_{\rm cal}$ and $0.0645$ for $P_{\rm PI}$. The independently evaluated normalized continuous directional sensitivities $\|D_{12}\|_{\Frob}^2/81$ are $0.020842$ and $0.004141$, respectively. These quantities are interaction diagnostics rather than gate error probabilities; together they show that the pulse-informed placement produces a substantially smaller residual interaction response for this processor.

\subsection{Held-out processor ensemble}

The pre-specified method was evaluated on 30 previously unused synthetic processor realizations. The median final simultaneous projected overlaps are
\begin{equation}
F_{\rm cal}^{\rm med}=0.984981,
\qquad
F_{\rm PI}^{\rm med}=0.993826.
\end{equation}
The corresponding means are $0.979738$ and $0.992850$, with sample standard deviations $0.014458$ and $0.003928$. Pulse-informed placement has the larger final $F_{\rm proj}$ in 28 realizations and selects the same physical placement in two. The mean paired gain is $0.013112$. A paired nonparametric bootstrap of the 30 processor-level gains, using $100000$ resamples with replacement and a percentile interval, gives a $95\%$ interval $[0.008994,0.018181]$. The median projected overlap deficit $1-F_{\rm proj}$ is reduced by $58.9\%$ relative to calibration-based placement.

\begin{figure*}[t]
\centering
\includegraphics[width=0.96\textwidth]{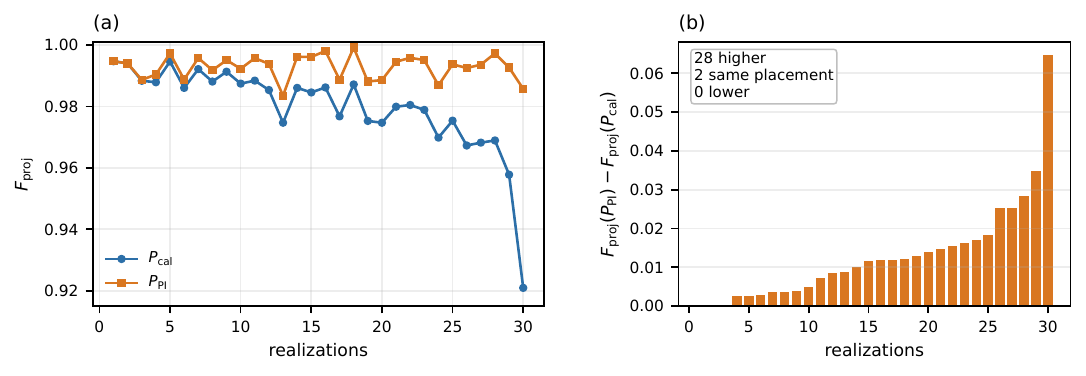}
\caption{\label{fig:ensemble}
Held-out 30-processor comparison. (a) Final projected overlaps for the calibration selected and pulse-informed placements, ordered only for visualization by their paired gain. (b) Processor by processor improvement $F_{\rm proj}(P_{\rm PI})-F_{\rm proj}(P_{\rm cal})$. Every point uses the same final pulse objective, budget, and direct coupled propagation.}
\end{figure*}

The median computational survivals are $0.993007$ for $P_{\rm cal}$ and $0.996734$ for $P_{\rm PI}$. The improvement in projected overlap is therefore not accompanied by a reduction in the median computational subspace survival. Figure~\ref{fig:ensemble} shows the paired final scores and processor by processor gains. A larger edge calibration budget sensitivity test is reported separately in Appendix~\ref{app:calsensitivity}.

\subsection{Search quality versus optimization effort}

The calibration baseline is intentionally inexpensive: after isolated-edge calibration it commits to one placement and therefore performs candidate level simultaneous quantum optimal control only for that selected location. Pulse-informed placement spends additional candidate control effort because it is designed to use simultaneous optimized dynamics during the placement decision. Its efficiency question is therefore not whether it is cheaper than calibration alone, but whether it can recover a near best control informed placement without giving the final control budget to every admissible location.

We evaluate that question on a separate processor realization by applying the common final budget $B$ to all 44 admissible placements. The finite budget exhaustive reference is
\begin{equation}
P_{\rm exh}^{(B)}=\arg\max_{P\in\calP_2}\mathcal F_B(P).
\label{eq:exhaustive}
\end{equation}
Because each inner solve uses the stated finite budget local optimizer and initialization, Eq.~\eqref{eq:exhaustive} is a placement search reference under that numerical protocol rather than a proof of the global continuous control optimum.

The best exhaustive reference score is $0.989957$. On the same processor, $P_{\rm PI}$ reaches $0.987817$ and ranks third among the 44 placements, while $P_{\rm cal}$ reaches $0.979015$ and ranks eighth. The corresponding gaps are
\begin{align}
\Delta_{\rm cal}&=0.010942,\\
\Delta_{\rm PI}&=0.002140,
\end{align}
so the pulse-informed choice reduces the gap to the best finite budget reference by $80.4\%$ relative to the calibration selected placement.

For this processor the initial pulse-informed candidate set contains 26 placements. Excluding the separate isolated edge calibration stage, the candidate control iteration counts are
\begin{align}
\mathcal I_{\rm cal}&=B=160,\\
\mathcal I_{\rm PI}&=26(40)+12(100)+4(160)=2880,\\
\mathcal I_{\rm exh}&=44(160)=7040.
\end{align}
Thus the calibration route is the cheapest because it does not conduct a control aware placement search. Pulse-informed placement occupies the intended intermediate regime: it uses more candidate level control effort than calibration alone, but $59.1\%$ fewer candidate control optimizer iterations than the exhaustive reference while recovering a placement close to the best finite budget score. The isolated edge calibration adds $12\times50=600$ two-qutrit iterations to both calibration guided approaches. Because a two-qutrit calibration iteration and a four-qutrit candidate control iteration have different computational costs, this mixed total is reported only as supplementary accounting rather than as an equal cost complexity metric.

Across the 30 held-out processors, the pulse-informed candidate stages use $2845.3$ candidate control optimizer iterations on average, $59.6\%$ below the $7040$-iteration exhaustive reference. These are optimizer iteration counts, not measured wall-clock or floating point operation speedups.

\begin{table}[b]
\caption{\label{tab:qualitycost}Search quality reference on the separate exhaustive evaluation processor. Candidate QOCT iterations exclude the isolated edge calibration overhead.}
\begin{ruledtabular}
\begin{tabular}{lccc}
Policy & Final $F_{\rm proj}$ & Rank / 44 & QOCT iter.\\
\hline
$P_{\rm cal}$ & 0.979015 & 8 & 160\\
$P_{\rm PI}$ & 0.987817 & 3 & 2880\\
Exhaustive reference & 0.989957 & 1 & 7040\\
\end{tabular}
\end{ruledtabular}
\end{table}

\begin{figure*}[t]
\centering
\includegraphics[width=0.96\textwidth]{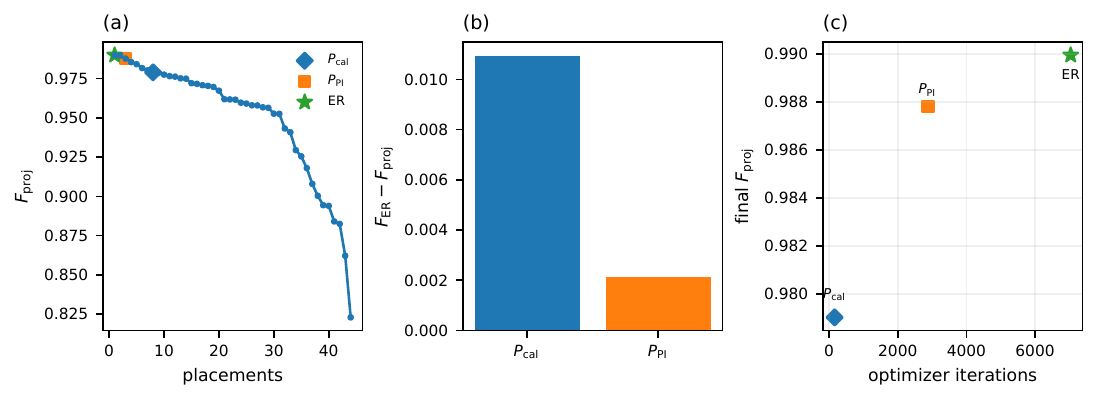}
\caption{\label{fig:qualitycost}
Search quality and candidate QOCT effort on the separate exhaustive reference processor. (a) Final projected overlap for all 44 admissible placements evaluated using the common final QOCT budget \(B\), together with the locations selected by the calibration based and pulse informed policies. (b) Gap of \(P_{\mathrm{cal}}\) and \(P_{\mathrm{PI}}\) to the best finite budget reference placement. Pulse-informed placement reduces this gap relative to calibration based placement. (c) Final \(F_{\mathrm{proj}}\) versus aggregate candidate QOCT optimizer iterations. \(P_{\mathrm{PI}}\) approaches the best finite budget reference with fewer candidate QOCT iterations than exhaustive final budget evaluation.}
\end{figure*}

As a secondary robustness diagnostic, the residual coupling scale is changed only after placement and pulse selection. Over three predetermined processors, the mean $F_{\rm proj}$ values at the nominal residual scale are $0.980152$ for $P_{\rm cal}$ and $0.992620$ for $P_{\rm PI}$; after doubling that scale they become $0.937923$ and $0.977692$. This post selection sweep shows that the placement difference remains relevant as residual coupling is strengthened, although it does not test selection under incorrectly estimated couplings.

\section{Generalization to larger parallel layers}

The placement formulation is not restricted to two simultaneous gates. Equations~\eqref{eq:placement}, \eqref{eq:admissible}, \eqref{eq:neighborhood}, and \eqref{eq:successive} are defined for arbitrary $L$: one enumerates or generates admissible $L$-edge placements, forms a calibration guided candidate region, evaluates candidate specific control responses, and progressively promotes the best candidates. In this sense the outer placement method is general and can be reused for larger parallel layers.

Two different scaling issues then appear. First, the number of legal placements grows rapidly with $L$ and the system's size. On a $5\times5$ nearest neighbor square grid, the number of disjoint $L$-edge matchings is 686 for $L=2$, 6632 for $L=3$, 39979 for $L=4$, and 157000 for $L=5$. Progressive pruning directly addresses this placement space growth by preventing most candidates from receiving the largest control budget.

Second, the cost of evaluating a candidate depends on how the simultaneous response is computed. The subsystem pair robustness terms in Eq.~\eqref{eq:loss} remain four-qutrit calculations with
\begin{equation}
d_{\rm pair}=3^4=81,
\end{equation}
and the number of pair terms grows only as
\begin{equation}
N_{\rm pair}=\binom{L}{2}=\frac{L(L-1)}{2}.
\end{equation}
This pair local structure follows the subsystem robust control construction, which replaces full system robustness calculations by constant size subsystem problems for local target operations \cite{Li2019,Yang2026}. By contrast, the exact direct candidate ranking metric used in the present headline study propagates all active qutrits. Extending that exact ranking unchanged to $L$ simultaneous two-transmon gates would use
\begin{equation}
d_{\rm full}=3^{2L},
\end{equation}
which becomes the dominant limitation.

The resulting scope is therefore precise. The \emph{placement formulation and progressive search logic} are generalized to $L>2$, and the \emph{robustness regularizer} is built from fixed size subsystem pairs. The present paper does not demonstrate an end to end scalable exact ranking implementation for arbitrary $L$. A practical larger $L$ realization can retain the same outer search but replace the global ranking at early stages by a validated blockwise, pairwise, tensor network, or hierarchical score, reserving higher cost verification for a small finalist set when feasible. Appendix~\ref{app:L3} reports a preliminary $L=3$ blockwise diagnostic illustrating one such direction, but it is not used as evidence for the main $L=2$ performance claims.

\begin{figure*}[t]
\centering
\includegraphics[width=0.94\textwidth]{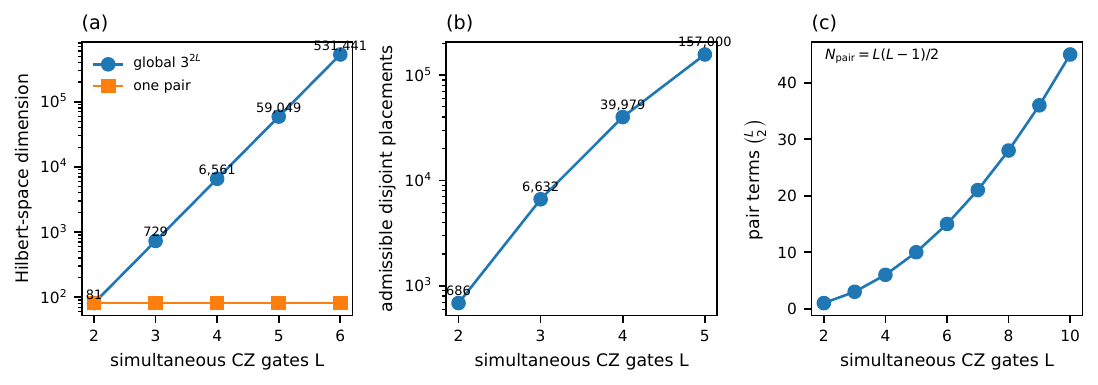}
\caption{\label{fig:scaling}
Scaling considerations for extending the placement framework to larger parallel layers. (a) Dimension of the global active qutrit space, \(d_{\mathrm{full}}=3^{2L}\), for \(L\) simultaneous two-qubit gates, compared with the fixed four qutrit dimension \(d_{\mathrm{pair}}=3^4=81\) of a single subsystem pair robustness calculation. (b) Number of admissible nonoverlapping \(L\) edge placements on a \(5\times5\) nearest neighbor square grid, illustrating the combinatorial growth of the physical placement search space. (c) Number of subsystem pair robustness terms, \(N_{\mathrm{pair}}=\binom{L}{2}=L(L-1)/2\). The pairwise robustness calculations therefore remain fixed in dimension, while their number grows quadratically with \(L\). In contrast, exact propagation of all active qutrits grows exponentially as \(3^{2L}\). The outer placement formulation is defined for general \(L\), whereas the present exact global candidate ranking validation is demonstrated only for \(L=2\).}
\end{figure*}

\section{Discussion}

The results support a simple physical distinction. Isolated edge calibration measures whether a native edge can realize a good CZ gate in its local subsystem, whereas physical placement asks whether a \emph{set} of such edges remains favorable after placement specific pulses and simultaneous coupled dynamics are included. These questions need not have the same ordering because changing $P$ changes the active frequencies, anharmonicities, intended couplings, residual interactions, and therefore the optimized control trajectory. The representative processor makes this distinction explicit: the calibration preferred placement is not the placement with the best final simultaneous response after candidate specific QOCT.

The same distinction clarifies the role of frequency crowding. We do not first sort candidates by a separate frequency spacing metric and we do not change the logical gate order. Rather, the frequencies and anharmonicities selected by $P$, together with the control excursions $u_P(t)$, determine the instantaneous spectral detunings inside $H_P(t)$. Their effect is therefore assessed dynamically by QOCT and by the final coupled propagation. In this sense the method performs re-ranking of a fixed parallel layer, not an independent frequency allocation or gate scheduling optimization.

The method therefore occupies a specific position relative to prior work. Calibration and crosstalk acquainted compilers choose physical resources using pre-existing device information \cite{Murali2019,Ding2020,CAMEL2025}; pulse-level approaches optimize controls or pulse/schedule combinations for a specified physical configuration \cite{Shi2019,AccQOC2020,Xie2022,Machnes2018}; and subsystem QOCT reduces the cost of robust pulse synthesis after a subsystem decomposition has been chosen \cite{Li2019,Yang2026}. The present outer loop treats admissible physical realizations themselves as competing control problems and feeds their post-QOCT simultaneous responses back into the placement state.

QMDM is the mechanism that keeps these two information levels connected. The static blocks $X_Q$, $G_{QQ}$, $X_E$, $B_{QE}$, $C_{EP}$, and $\calT$ describe what physical system corresponds to a candidate placement. The dynamic block $X_P^{(b)}$ records what happens \emph{after that candidate has been controlled}. The proposed method therefore does not infer $P_{\rm PI}$ from a static crosstalk score. Instead, it repeatedly maps $P$ to $H_P$, solves the candidate control problem, measures the directly propagated simultaneous response, and feeds that information back into the placement state. This feedback is the central theoretical difference between calibration-based and pulse-informed placement.

The representative interaction diagnostics also show why the effect should not be interpreted as simply choosing the smallest bare residual coupling. For $P_{\rm PI}$ the final interaction picture deviation and continuous first-order sensitivity are both substantially smaller than for $P_{\rm cal}$, but the optimized response depends on the full placement specific control trajectory. Pulse-informed placement is therefore control aware: the residual interaction is assessed after it has been dressed by the pulse designed for the candidate Hamiltonian.

The search effort comparison has a similarly limited interpretation. $P_{\rm cal}$ is cheaper because it makes no control based search over placement space. The exhaustive finite budget reference is control aware but assigns the final candidate QOCT budget to every admissible placement. $P_{\rm PI}$ intentionally occupies the intermediate regime: it spends more control effort than calibration only selection, but progressively removes candidates so that most placements never receive the largest budget. The reported optimizer iteration reductions quantify that allocation under the $L=2$ candidate model; they are not wall-clock speedups.

Several limitations remain. The processor ensemble is synthetic, the nominal residual couplings supplied to QMDM are assumed known, the dynamics are closed and coherent, and the main exact verification propagates only the four active transmons of the $L=2$ layer. Transfer functions, decoherence, calibration drift, active spectators, and uncertainty in the QMDM processor blocks are not included. The candidate QOCT is also nonconvex and uses one pre-specified deterministic initialization, so systematic multi-start placement ranking remains an important robustness test. An equal candidate budget ablation, for example applying the final QOCT budget to a fixed number of top calibration candidates, would further separate the value of progressive allocation from the value of evaluating multiple placements. Finally, the outer QMDM formulation and pulse-informed feedback are general in $L$, but the exact candidate ranking metric used in the main study would scale with the global active Hilbert space if extended unchanged. A practical large $L$ realization therefore requires a separately validated blockwise, pairwise, or hierarchical ranking and verification strategy.

\section{Conclusion}

We introduced a QMDM guided pulse-informed physical placement framework for parallel CZ gates. The central idea is to make the placement decision depend on the pulse control problem that the placement itself creates. QMDM stores the processor, edge, coupling, incidence, and task information needed to instantiate a candidate Hamiltonian $H_P$ and, after candidate specific QOCT, is updated with the directly propagated simultaneous response. The outer search then uses this pulse-conditioned candidate state to rank and promote physical realizations. In this way, the information flow is closed from placement to pulse and back to placement.

This construction makes the comparison with calibration based placement precise. $P_{\rm cal}$ is fixed from isolated-edge qualities before simultaneous optimized dynamics are used, whereas $P_{\rm PI}$ is selected after several candidates have been evaluated through their own optimized simultaneous pulses. Both selected placements are judged with the same final candidate QOCT formulation and coupled system metric. On 30 held-out synthetic $3\times3$ processors, the median $F_{\rm proj}$ is $0.984981$ for $P_{\rm cal}$ and $0.993826$ for $P_{\rm PI}$, with the pulse-informed result higher in 28 cases and the same physical placement selected in two. On the separate 44-placement exhaustive reference processor, $P_{\rm PI}$ ranks third and reduces the gap to the best finite-budget reference placement while requiring substantially fewer candidate QOCT iterations than exhaustive final budget evaluation.

The supported conclusion is therefore not that isolated calibration is unnecessary. Calibration provides an efficient prior over the placement space. The result is that calibration alone does not fully determine the best simultaneous physical realization: candidate specific optimized pulse dynamics can re-rank physically admissible placements, and QMDM provides a structured way to feed that information back into the outer placement decision. The framework is defined for general $L$; the present exact numerical ranking is validated for $L=2$, and a scalable larger $L$ ranking/verification layer remains future work.

\appendix

\section{Directional integral and truncation checks}
\label{app:validation}

The directional derivative in Eq.~\eqref{eq:Dkj} is the standard first-order toggling frame response to an inter-subsystem perturbation used in subsystem robust control \cite{Li2019,Yang2026}. For a piecewise constant pair-local crosstalk free Hamiltonian $H_{0,s}^{(kj)}$ on segment $s$, write
\begin{equation}
H_{0,s}^{(kj)}=V_s\operatorname{diag}(E_{s,1},\ldots,E_{s,d})V_s^\dagger
\end{equation}
and define $\widehat H_{x,s}=V_s^\dagger H_{S_kS_j}V_s$. With the normalized convention $\operatorname{sinc}(x)=\sin(\pi x)/(\pi x)$, the exact segment contribution to the toggling frame integral is
\begin{equation}
I_s=V_s(\widehat H_{x,s}\odot\Gamma_s)V_s^\dagger,
\end{equation}
where
\begin{equation}
(\Gamma_s)_{mn}=\Delta t\,
e^{i(E_{s,m}-E_{s,n})\Delta t/2}
\operatorname{sinc}\!\left[\frac{(E_{s,m}-E_{s,n})\Delta t}{2\pi}\right].
\end{equation}
If $U_{0,kj}(t_s)$ is the pair local propagator accumulated before segment $s$,
\begin{equation}
I_{kj}=\sum_sU_{0,kj}^\dagger(t_s)I_sU_{0,kj}(t_s),
\end{equation}
and $\|D_{kj}\|_{\Frob}=\|I_{kj}\|_{\Frob}$.

The same integral can alternatively be evaluated with block matrix exponential constructions of the Van Loan type \cite{VanLoan1978,Li2019}. The benchmark optimizer instead uses the cheaper boundary sampled approximation
\begin{align}
K_s&=U_{0,kj}^\dagger(t_s)H_{S_kS_j}U_{0,kj}(t_s),\\
\widetilde I_{kj}&=\Delta t\left[
\frac{K_0+K_{N_s}}{2}+\sum_{s=1}^{N_s-1}K_s\right],\\
\widetilde r_{kj}&=\frac{\|\widetilde I_{kj}\|_{\Frob}^2}{d_{kj}}.
\end{align}
For the representative processor, the exact normalized continuous sensitivities are $0.020842$ for $P_{\rm cal}$ and $0.004141$ for $P_{\rm PI}$, whereas the corresponding boundary proxy values are $0.057895$ and $0.023896$. The proxy is therefore used only as a regularization signal; all placement rankings and headline gate scores use direct coupled propagation.

\begin{figure}[t]
\centering
\includegraphics[width=\columnwidth]{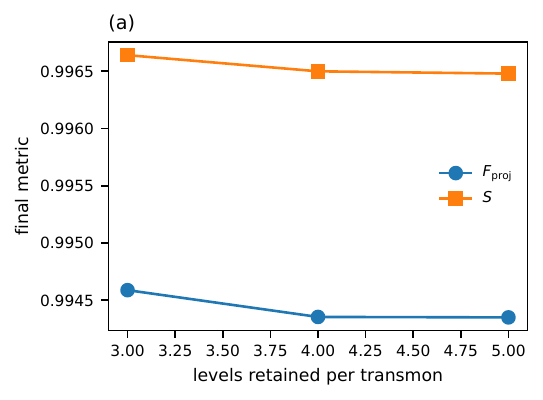}
\caption{\label{fig:validation}
The representative pulse selected by the pulse-informed search is reevaluated with three, four, and five retained levels per transmon. The small change between the four and five level results indicates convergence at the truncation used for the main benchmark.}
\end{figure}

The qutrit truncation was checked by reevaluating the representative pulse with three, four, and five levels per transmon. The resulting projected overlaps are $0.9945883$, $0.9943543$, and $0.9943503$, respectively; the four and five level values differ by $4.0\times10^{-6}$. Figure~\ref{fig:validation} summarizes the derivative and truncation checks.

\section{Exploratory $L=3$ blockwise diagnostic}
\label{app:L3}

To examine whether the subsystem-based control ingredients can be used beyond the \(L=2\) benchmark, we performed an exploratory blockwise calculation for \(L=3\), corresponding to three simultaneous controlled-\(Z\) subsystems. Three development processor realizations were considered. For each candidate placement \(P\), the three subsystem pairs \((1,2)\), \((1,3)\), and \((2,3)\) were evaluated separately by direct four qutrit propagation. We then defined the candidate level diagnostic as the worst pairwise projected overlap,
$$F_{\text{worst}}(P) = \min_{k < j} F_{kj}(P)$$
This choice asks whether a placement remains acceptable for its most strongly affected subsystem pair, rather than averaging over the three pairs.

\begin{figure}[t]
\centering
\includegraphics[width=\columnwidth]{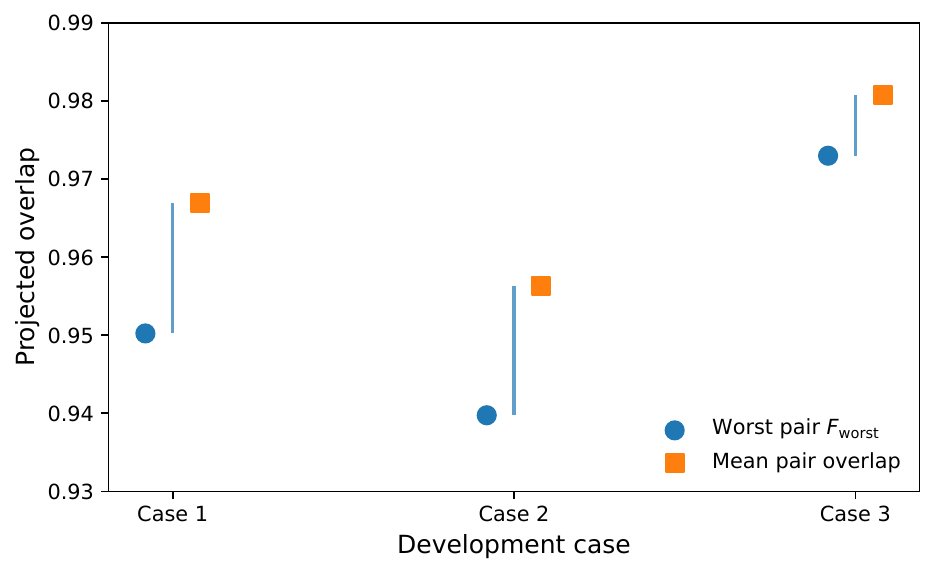}
\caption{\label{fig:L3}
Exploratory $L=3$ blockwise diagnostic. (a) Worst directly propagated subsystem pair overlap for the compared development choices. (b) Source of the retained candidate in the two-candidate rule.}
\end{figure}

Figure \ref{fig:L3} summarizes this exploratory calculation. The calculation is performed for three development processor realizations to verify that the fixed size pairwise diagnostic can be evaluated when more than two gate subsystems are active. The purpose of this test is only to illustrate the \(L=3\) pairwise construction. It is not a global six qutrit fidelity calculation and does not implement the complete progressive pulse-informed placement search for \(L=3\).

\section{Calibration-budget sensitivity}
\label{app:calsensitivity}

The main calibration baseline uses eight pulse segments and 50 local Adam iterations per native edge. To test sensitivity to the numerical accuracy of these isolated calibration values, we repeat the same placement rule with 12 pulse segments and 160 local iterations per edge. In both cases the selected placement is subsequently evaluated with the final candidate control protocol used for $P_{\rm PI}$.

Increasing the edge calibration effort raises the median final $F_{\rm proj}(P_{\rm cal})$ from $0.984981$ to $0.988372$. Relative to the stronger calibration placement, $P_{\rm PI}$ is higher in 22 processors, the same placement is selected in three, and the stronger calibration placement is higher in five. Thus the pulse-informed method is not pointwise dominant, and part of its advantage over the main baseline is reduced when isolated-edge calibration is made more accurate. One stronger calibration realization also produces a severe low simultaneous score, $F_{\rm proj}=0.091902$, despite a high isolated calibration product; with a single deterministic control initialization, the present data do not distinguish an intrinsically difficult simultaneous placement from local optimizer trapping in that case.

\begin{table*}[t]
\caption{\label{tab:calsensitivity}Sensitivity of the calibration based placement to the isolated edge optimization budget. Both rows use the same final candidate control evaluation after the placement has been selected.}
\begin{ruledtabular}
\begin{tabular}{lcccc}
Edge calibration calculation & Median final $F_{\rm proj}(P_{\rm cal})$ & $P_{\rm PI}$ higher & Same placement & $P_{\rm cal}$ higher\\
\hline
8 segments, 50 iterations & 0.984981 & 28 & 2 & 0\\
12 segments, 160 iterations & 0.988372 & 22 & 3 & 5\\
\end{tabular}
\end{ruledtabular}
\end{table*}


%

\end{document}